%% file: main.tex
\documentclass[runningheads]{llncs}
\usepackage[T1]{fontenc}
\usepackage{graphicx,verbatim}
\usepackage{booktabs}
\usepackage{multirow}  % tables
\usepackage{stmaryrd} % arrows
\usepackage{amssymb}
\usepackage{amsmath}
\usepackage{xcolor}
\usepackage{colortbl}
\usepackage[colorlinks=true,
            linkcolor=blue,     % For internal links (sections, tables, etc.)
            citecolor=blue,     % For citation links
            urlcolor=blue%For % For hyperlinks (URLs)
           ]{hyperref}
\begin{document}
%
% \title{Contribution Title}
\title{\emph{MoPET}: Parameter-Efficient Mixture-of-Experts for Unified Medical Image Classification}
\titlerunning{Parameter-Efficient MoE for Unified Medical Image Classification}
\author{Sebastian Doerrich\thanks{These authors contributed equally to this work.} \and 
Daniel Würtinger\protect\footnotemark[1] \and
Francesco Di Salvo \and
Shyam Nandan Rai \and
Christian Ledig}
\authorrunning{S. Doerrich et al.}
\institute{xAILab Bamberg, University of Bamberg, Bamberg, Germany
\email{sebastian.doerrich@uni-bamberg.de}}
\maketitle              % typeset the header of the contribution
\begin{abstract}
Adapting deep learning models to profound clinical heterogeneity typically relies on parameter-efficient fine-tuning (PEFT) to avoid the severe overfitting associated with full end-to-end network updates. Although PEFT successfully navigates limited data scenarios, it inherently forces the training of a separate, isolated adapter for every specific diagnostic task. Consolidating these isolated adapters into a single generalist network risks negative transfer, as optimization gradients from conflicting visual domains interfere.
To address this, we propose \emph{MoPET}, a mixture-of-experts (MoE) method that uses a learned sparse router to direct each input through a small subset of low-rank PEFT experts injected into a frozen foundation model, sharing capacity across datasets while limiting cross-domain gradient conflict. Through selected evaluations on the MedMNIST benchmark, we first establish that PEFT outperforms full network updates, improving average accuracy from 86.50\% to 88.97\%. We then show that a single \emph{MoPET} model consolidates four heterogeneous datasets into one network, improving average accuracy over the best isolated PEFT adapters (93.46\% versus 92.83\%). Finally, we show that co-training with auxiliary datasets improves accuracy on data-constrained clinical targets, raising average target accuracy over the strongest isolated adapter from 81.58\% to 83.58\%. Our source code is publicly available at \url{https://github.com/sdoerrich97/mopet}.

\end{abstract}
\section{Introduction}
\label{sec:introduction}
\begin{figure}[htb]
\centering
\includegraphics[width=0.9\textwidth]{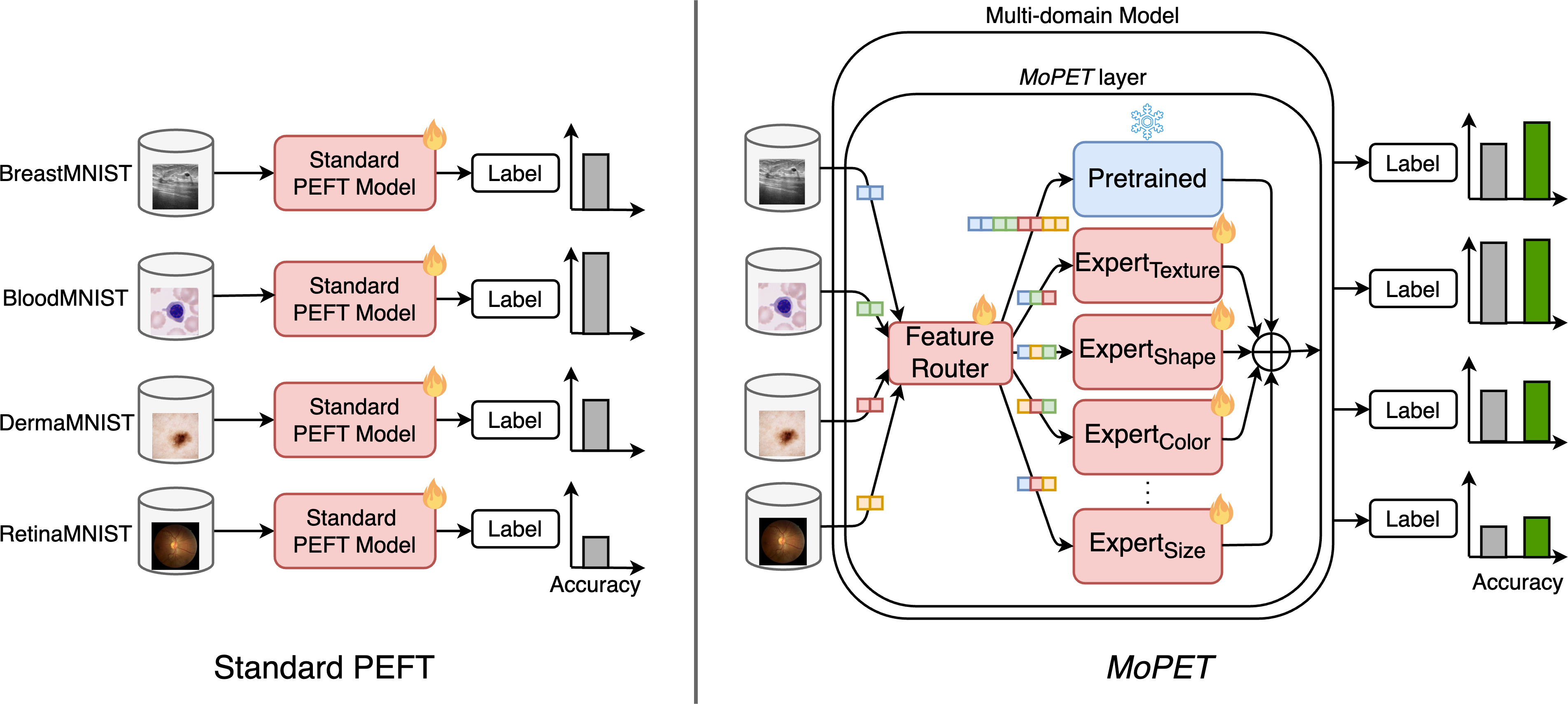}
\caption{Left: Standard Parameter-Efficient Fine-Tuning (PEFT) requires independent, disconnected models for distinct medical modalities. Right: \emph{MoPET} dynamically routes diverse anatomical inputs through specialized low-rank experts within a single unified architecture, preventing gradient collision and boosting overall accuracy.}
\label{fig:pull_figure}
\end{figure}

The transition from curated datasets to real-world clinical deployment requires models capable of reasoning across profound anatomical heterogeneity. Generalist foundation models trained on natural images, such as CLIP~\cite{Radford2021} and DINOv3~\cite{Simeoni2025}, provide highly robust visual representations. However, adapting these massive architectures to the nuanced visual domains of specific medical tasks presents a structural dilemma~\cite{Tiu2022}. Conversely, medical-specific foundation models~\cite{Chen2024c,Wang2022,Zhang2025a} capture domain-relevant features but typically specialize in narrow anatomical regions, demanding necessary adaptation for broader clinical application. Regardless of the base model, full end-to-end finetuning remains computationally prohibitive and consistently overfits in limited data scenarios. Parameter-efficient techniques like Low-Rank Adaptation (LoRA)~\cite{Hu2021} resolve this computational burden and data scarcity problem by freezing pretrained weights and updating only injected low-rank matrices, requiring substantially fewer training examples. Although LoRA successfully retains generalization, it requires training a separate adapter for every new task or modality. This fragmented strategy spawns a massive proliferation of disparate adapters, which complicates deployment in resource-constrained clinical settings.

To unify these separate adapters into a single network, architectures must overcome the common problem of negative transfer~\cite{Ruder2017}. When optimization gradients from diverse visual domains conflict (such as the high-frequency cellular textures of histopathology versus the low-frequency geometric patterns of ultrasound) model training becomes challenging. Mixture of Experts (MoE) architectures~\cite{Dai2024,Jacobs1991}, including the recent M4oE~\cite{Jiang2024}, MedMoE~\cite{Chopra2025}, and Med-MoE~\cite{Jiang2024Med_MoE}, resolve this interference through sparse routing mechanisms that dynamically assign features to modality-specific experts. Closest to our setting, mixtures of low-rank experts couple sparse routing with parameter-efficient adapters, though primarily in language models and with a homogeneous expert pool~\cite{Zadouri2024,Komatsuzaki2023}.
Building on these principles, we introduce \emph{MoPET}, a Mixture of Parameter-Efficient Fine-Tuned Experts that condenses distinct medical image classification tasks into a single model. Unlike prior medical MoE frameworks that tie experts to predefined imaging modalities, \emph{MoPET} learns its expert assignments from data, without modality labels at the routing stage. A learned router directs each input through a small subset of a heterogeneous pool of LoRA and BOFT experts, so that domain-specific adaptation and shared semantic reasoning need not compete for the same parameters, while a dynamic minimum-size sampler and per-dataset classification heads address the scale imbalance and disjoint label spaces across datasets. This reduces gradient conflict across domains while retaining the memory efficiency of low-rank adaptation in a single unified model (\figurename~\ref{fig:pull_figure}).
Our contributions are as follows:
\begin{itemize}
    \item We first establish across 12 distinct medical datasets that parameter-efficient finetuning outperforms full end-to-end network updates. This motivates our design choice of building the unified architecture from low-rank adapters rather than full updates.
    \item We introduce \emph{MoPET}, a novel Mixture of Experts framework built entirely from parameter-efficient modules that condenses distinct medical image classification tasks into a single model. We demonstrate that this outperforms isolated parameter-efficient adapters trained on individual domains for a selected set of four heterogeneous datasets of varying sizes and modalities.
    \item Finally, we reveal a novel cross-domain training dynamic where the inclusion of auxiliary medical datasets acts as a performance booster. We demonstrate this by elevating the predictive accuracy across three distinct clinical domains through joint training with auxiliary data pools.
\end{itemize}
\section{Methodology}
\begin{figure}[htb]
\centering
\includegraphics[width=0.8\linewidth]{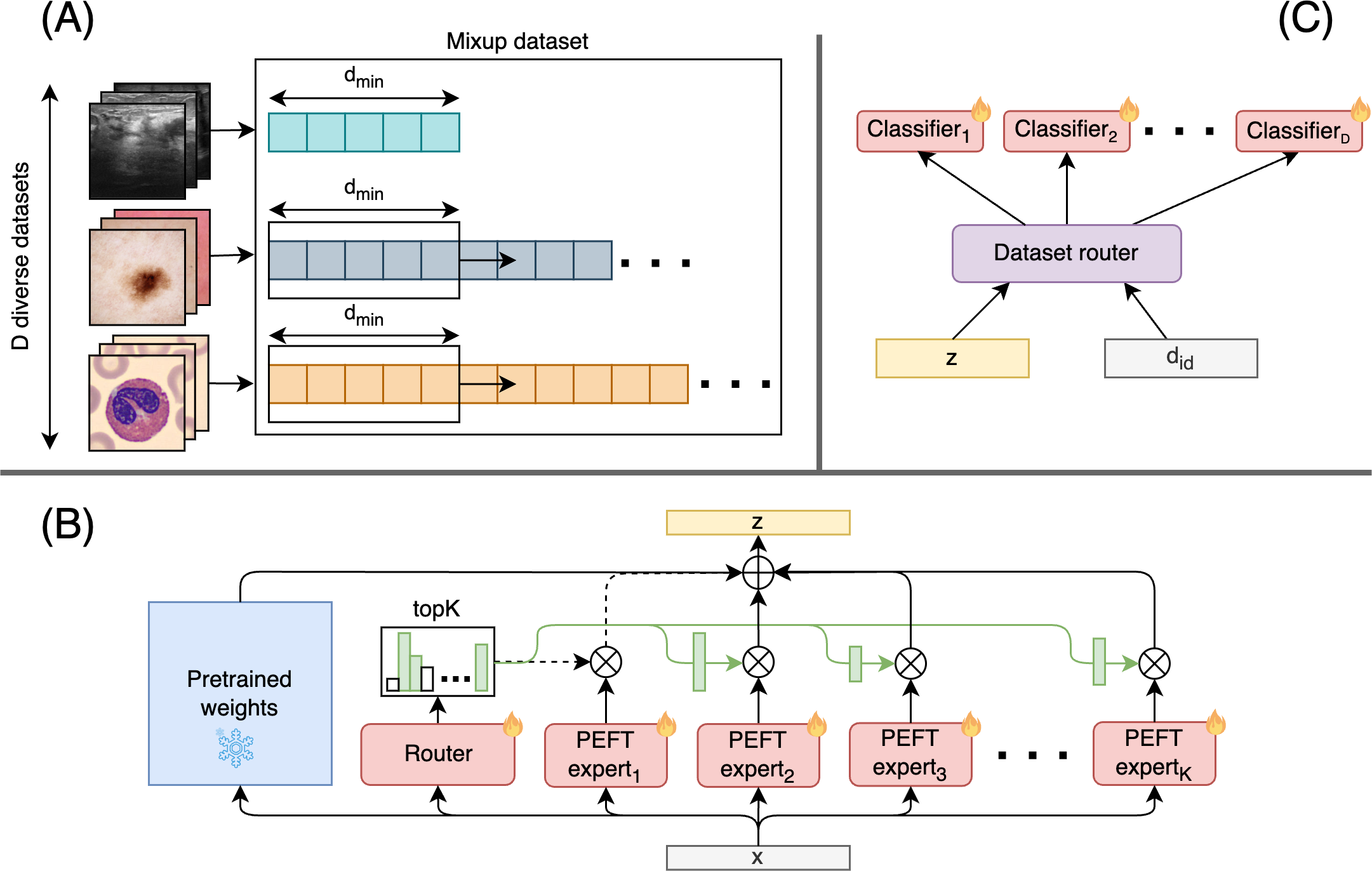}
\caption{Overview of \emph{MoPET}. First, a dynamic sampling protocol extracts balanced batches of size $d_{\min}$ from each dataset (A). Second, during the forward pass, an internal router activates the top-$k$ parameter-efficient (PEFT) experts to process the input representation $\mathbf{x}$ alongside the frozen pretrained weights, producing the updated token representation $\mathbf{z}$ (B). Finally, a dataset router utilizes the domain identifier $d_{id}$ to direct the terminal representation $\mathbf{z}$ to the corresponding task-specific classifier (C).}
\label{fig:methodology}
\end{figure}

We propose \emph{MoPET}, an architecture that adapts the mixture-of-experts paradigm to parameter-efficient finetuning, for establishing a unified classification model that simultaneously reasons across multiple distinct medical datasets.
When trained jointly, diverse medical modalities typically suffer from negative transfer, thereby degrading overall predictive performance.
To resolve this, our method integrates 4 sequential components: a dynamic sampling strategy to balance data ingestion, a routing mechanism to isolate conflicting features into specialized low-rank pathways, a load-balancing loss to ensure optimal expert utilization, and task-specific classification heads (\figurename~\ref{fig:methodology}).

\subsection{Feature Aware Routing Architecture}
Standard end-to-end finetuning of large scale networks on medical data frequently leads to overfitting and catastrophic forgetting. To circumvent this, \emph{MoPET} integrates the Mixture of Experts (MoE) paradigm with Parameter Efficient Fine Tuning (PEFT). We maintain a frozen pretrained backbone and inject a small set of trainable low rank matrices (experts) into the query, key, and value projection layers of the self-attention blocks to form a unified adaptation layer. This approach isolates the learning of new domain specific features while preserving the robust generalist representations of the original network.
However, simultaneously updating a single set of shared adapters across divergent clinical tasks can cause negative transfer. To physically separate competing gradient updates among the $K$ introduced experts, we deploy a learnable gating network $R$. For a given input token vector $\mathbf{x}$, the gating network $R(\mathbf{x})$ computes raw, unnormalized routing scores to evaluate the relevance of each available expert. We define the final hidden representation $\mathbf{z}$ by augmenting the frozen weight matrix $\mathbf{W}$ with the dynamically weighted expert outputs $\mathbf{e}_i(\mathbf{x})$:
\begin{equation}
    \mathbf{z} = \mathbf{W}\mathbf{x} + \sum_{i = 1}^{K} \text{softmax}(\text{TopK}(R(\mathbf{x}), k))_i \mathbf{e}_i(\mathbf{x})
\end{equation}
The $\text{TopK}$ operation retains the $k$ highest routing scores and sets the rest to $-\infty$, so the softmax assigns zero weight to unselected experts and the sum reduces to the $k$ active ones. Each expert $\mathbf{e}_i$ is a parameter-efficient module, either a LoRA or a BOFT adapter, while the frozen projection $\mathbf{W}\mathbf{x}$ acts as an always-on shared path. Routing each token through a subset of experts lets the model adapt to divergent domains while limiting interference between their gradient updates.

To encourage the router to use the full expert pool and avoid expert collapse, where the router repeatedly selects a narrow subset of experts, we add the differentiable load-balancing loss of DeepSeekMoE~\cite{Dai2024}:
\begin{equation}
\mathcal{L}_{\text{load}} = K\sum_{i=1}^{K} f_i\,p_i,
\end{equation}
Here, $f_i$ is the fraction of tokens routed to expert $i$, and $p_i$ is its mean routing probability over the batch. Minimizing this loss promotes a uniform distribution of tokens across all $K$ experts.

We formulate our composite training set as $\mathcal{D} = \{(\mathbf{X}_d, \mathbf{Y}_d)\}_{d=1}^D$, where $\mathbf{X}_d$ denotes the input images, $\mathbf{Y}_d$ the corresponding labels, and $D$ indicates the total number of distinct clinical domains. Because this composite dataset $\mathcal{D}$ contains completely divergent label spaces, we attach $D$ independent linear classification heads to the frozen backbone. While the expert layers dynamically route and mix features internally, a dataset router explicitly uses the domain identifier $d_{id}$ to map the shared terminal representations $\mathbf{z}$ strictly into their corresponding semantic spaces, enabling dedicated predictions for every dataset $d \in \{1, \dots, D\}$.
We train the experts, router, and heads jointly by minimizing $\mathcal{L} = \mathcal{L}_{\text{CE}} + \lambda\,\mathcal{L}_{\text{load}}$, where $\mathcal{L}_{\text{CE}}$ is the cross-entropy on each sample's dataset-specific head.

\subsection{Dynamic Multi Domain Sampling}
Medical imaging benchmarks frequently exhibit highly skewed class distributions and severe volumetric disparities. If we sample naively, large datasets overwhelm the network and starve smaller domains. To enforce stable optimization across the composite training set $\mathcal{D}$, we deploy a dynamic sliding window sampling protocol. We structure each training epoch to contain exactly $D \times d_{\min}$ samples, where $d_{\min}$ is the number of training samples in the smallest domain.
This protocol guarantees the model processes the entirety of the smallest datasets in every single epoch while progressively iterating through the unseen samples of the larger datasets over successive epochs, strictly preventing domain starvation.
\subsection{Implementation Details}
We instantiate \emph{MoPET} using the DINOv3~\cite{Simeoni2025} Base architecture from the timm library~\cite{rw2019timm} as our frozen backbone, which processes inputs at a spatial resolution of $256 \times 256$ pixels. While model performance typically scales with expert count~\cite{Pan2026}, we constrain \emph{MoPET}'s complexity to $K=32$ experts per Transformer block following established scaling conventions~\cite{OpenAI2025}, while utilizing the pre-trained weight matrix as a shared expert~\cite{Dai2024}. For the specialized pathways, we use 20 Low-Rank Adaptation (LoRA) experts with a rank $r=8$ and a scaling factor $\alpha=8$, alongside 12 BOFT (Butterfly Orthogonal Fine-Tuning) experts utilizing a block size of $8$ and a butterfly factor of $1$. During isolated parameter-efficient finetuning, LoRA achieves the highest overall accuracy, while BOFT surpasses it on a subset of datasets (notably Breast and Retina). This complementarity motivates our hybrid expert pool, with more experts allocated to the stronger LoRA family. A sparse gating mechanism activates the top-$k=12$ experts per forward pass. In total, \emph{MoPET} trains 7.4M parameters ($\approx$8.7\% of the 86M backbone). While a single isolated adapter is lighter (LoRA 0.30M, BOFT 0.10M), it serves only one task; \emph{MoPET} instead unifies all tasks in one model, replacing a per-task collection of adapters and the model zoo it entails. Finally, we attach task-specific classification heads for each distinct dataset.
We conduct training for $75$ epochs with a batch size of $128$ using AdamW~\cite{loshchilov2019decoupledweightdecayregularization} (initial learning rate of $1 \times 10^{-3}$), cosine annealing scheduling~\cite{loshchilov2017sgdrstochasticgradientdescent}, early stopping based on validation performance (patience of 10 epochs) and a loss weight $\lambda = 0.05$.
\section{Experiments and Results}
We evaluate on twelve 2D datasets from the MedMNIST+ collection~\cite{Yang2023,yang2024medmnistplus} (2--11 classes; CC BY 4.0 / CC BY-NC 4.0), provided at $224\times224$ pixels and resized to $256\times256$ with bicubic interpolation to match the backbone. We follow the official data splits throughout. They cover eight imaging modalities across a wide range of scales: blood cell microscopy (Blood~\cite{Acevedo2020}), breast ultrasound (Breast~\cite{Al-Dhabyani2020}), chest X-ray (Chest~\cite{Wang2017ChestXRay8HC}, Pneumonia~\cite{Kermany2018}), dermatoscopy (Derma~\cite{Tschandl2018,Codella2019}), retinal OCT (OCT~\cite{Kermany2018}), abdominal CT in axial, coronal, and sagittal views of the same volumes (OrganA, OrganC, OrganS~\cite{BILIC2023102680,Xu2019}, treated as distinct tasks), colon pathology (Path~\cite{Kather2019}), kidney cortex microscopy (Tissue~\cite{Yang2023}), and fundus photography (Retina~\cite{Liu2022}). Training sets span from 546 images (Breast) to roughly 165{,}000 (Tissue), a scale heterogeneity that motivates our dynamic multi-domain sampler.

\subsection{Isolating Adaptation Strategy from Backbone Influence}
\label{sec:comparative_analysis}
First, we establish that parameter-efficient fine-tuning outperforms full end-to-end network updates. To isolate the influence of the backbone from the fine-tuning method, we compare CLIP~\cite{Radford2021}, DINO~\cite{Caron2021}, and DINOv3~\cite{Simeoni2025} under full end-to-end fine-tuning and the PEFT techniques LoRA~\cite{Hu2021}, BOFT~\cite{liu2024boft}, FourierFT~\cite{Gao2024FourierFT}, and AdaptFormer~\cite{chen2022adaptformer}. \tablename~\ref{tab:results1} reports the mean accuracy, evaluated at an operating point of 0.5, across the test splits of all twelve datasets and three seed runs. The end-to-end results for DINO and CLIP, as well as the DINOv3 training procedure, are adapted from~\cite{Doerrich2025}. For PEFT, we adjusted the learning rate from 0.0001 to 0.001. 
When evaluating the backbones under parameter efficient settings, the architectures exhibit high performance stability. A Friedman test reveals no statistically significant differences among the foundation models ($\chi^2 = 3.63$, $p = 0.163$, $\alpha = 0.05$). While a non-significant test does not establish equivalence, backbone selection alone does not provide reliable performance separation in clinical tasks.

Conversely, the method of parameter update dictates diagnostic success. On aggregate across the evaluated datasets, PEFT methods, specifically LoRA and BOFT, outperform traditional full network updates. Pairwise Wilcoxon signed-rank tests with Bonferroni correction confirm this: both LoRA and BOFT significantly outperform end-to-end fine-tuning ($p \le 1.2 \times 10^{-3}$) and AdaptFormer ($p < 10^{-8}$). LoRA additionally outperforms FourierFT ($p = 4.7 \times 10^{-7}$), and BOFT likewise outperforms FourierFT ($p = 3.7 \times 10^{-5}$), whereas LoRA and BOFT are statistically indistinguishable ($p = 0.08$) and end-to-end fine-tuning and FourierFT do not differ significantly ($p = 1.00$).

\begin{table}[htb]
\caption{Average test accuracy in \%, evaluated at an operating point of 0.5, of different PEFT methods and backbones across all twelve MedMNIST datasets and three seed runs in comparison to end-to-end fine-tuning. Best backbone per fine-tuning method is shown in bold (columns), while best fine-tuning method per backbone is underlined (rows).}
\centering
\setlength{\tabcolsep}{6.5pt}
\begin{tabular}{lccccc}
\toprule
Model & LoRA & BOFT & FourierFT & AdaptFormer & endToEnd \\
\midrule
CLIP   & \underline{$88.04 {\scriptstyle \pm 0.87}$}
       & $87.37 {\scriptstyle \pm 0.92}$
       & $84.57 {\scriptstyle \pm 0.61}$
       & $\mathbf{82.84 {\scriptstyle \pm 1.74}}$
       & $82.75 {\scriptstyle \pm 1.01}$ \\
DINO   & \underline{$88.14 {\scriptstyle \pm 0.81}$}
       & $87.28 {\scriptstyle \pm 1.19}$
       & $85.43 {\scriptstyle \pm 0.60}$
       & $82.08 {\scriptstyle \pm 0.96}$
       & $84.84 {\scriptstyle \pm 1.07}$ \\
DINOv3 & \underline{$\mathbf{88.97 {\scriptstyle \pm 0.73}}$}
       & $\mathbf{88.62 {\scriptstyle \pm 0.68}}$
       & $\mathbf{86.72 {\scriptstyle \pm 0.21}}$
       & $79.82 {\scriptstyle \pm 1.13}$
       & $\mathbf{86.50 {\scriptstyle \pm 1.15}}$ \\
\bottomrule
\end{tabular}
\label{tab:results1}
\end{table}

\subsection{Cross-Dataset Adaptation}
Next, we evaluate the cross-dataset adaptation capabilities of \emph{MoPET}. For this, we train a single, unified model simultaneously on a selected pool of four heterogeneous datasets of varying sizes and modalities (BloodMNIST, BreastMNIST, DermaMNIST, and PathMNIST) and compare it against the individually trained baseline adapters using DINOv3 as backbone. In this multi-domain configuration, our dynamic sampling protocol restricts the number of samples per dataset per epoch to $d_{\min} = 546$, matching the training size of the smallest domain (BreastMNIST). This mechanism prevents the larger datasets from dominating the gradient updates, thereby ensuring balanced anatomical representation.
\tablename~\ref{tab:results2} reports test accuracy across the four test splits for the individually trained baselines and our single \emph{MoPET} model, as the mean over three seeds. Compared to the strongest isolated PEFT adapter, \emph{MoPET} improves on BreastMNIST, DermaMNIST, and PathMNIST and stays within 0.11\% on BloodMNIST, raising the average from 92.83\% (BOFT) to 93.46\%. End-to-end fine-tuning remains strongest on PathMNIST but trails on the smaller, more heterogeneous targets. That a single model matches or exceeds the best per-dataset adapter on three of four datasets indicates that the routed experts share features across domains rather than overfitting to dataset-specific cues.

\begin{table}[htb]
\caption{Test accuracy in \%, evaluated at an operating point of 0.5, comparing traditional fine-tuning and isolated parameter-efficient adapters against the unified \emph{MoPET} framework using the DINOv3 backbone, on a selected pool of four heterogeneous datasets of varying sizes and modalities (BloodMNIST, BreastMNIST, DermaMNIST, and PathMNIST). All values are the mean over three random seeds. The best method per dataset is underlined; \emph{MoPET} is shown in bold where it ranks among the top three methods.}
\centering
\setlength{\tabcolsep}{3.5pt}
\begin{tabular}{lccccc}
\toprule
Method
 & BloodMNIST 
 & BreastMNIST 
 & DermaMNIST 
 & PathMNIST 
 & Average \\
 \midrule
End-to-End
& 98.49
& 85.26
& 81.23
& \underline{96.13}
& 90.28 \\
Linear Probing
& 97.86
& 86.54
& 82.28
& 93.90
& 90.15 \\
\midrule
AdaptFormer
& 96.50
& 73.29
& 69.89
& 89.06
& 82.19 \\
BOFT
& \underline{98.81}
& 90.38
& 86.93
& 95.21
& 92.83 \\
FourierFT
& 98.40
& 88.46
& 83.18
& 94.52
& 91.14 \\
LoRA
& 98.75
& 88.25
& 87.43
& 95.07
& 92.38 \\
\midrule
\emph{MoPET}
& \textbf{98.70}
& \underline{\textbf{90.81}}
& \underline{\textbf{88.46}}
& \textbf{95.87}
& \underline{\textbf{93.46}} \\
\bottomrule
\end{tabular}
\label{tab:results2}
\end{table}

\begin{table}[htb]
\caption{Test accuracy in \% between traditional fine-tuning, isolated parameter-efficient adapters, and the \emph{MoPET} boosting setup using the DINOv3 backbone. \emph{MoPET} boosts each primary target (BreastMNIST, RetinaMNIST, DermaMNIST) with a different, hand-selected pool of auxiliary datasets during training. All values are the mean over three random seeds; the best method per target is underlined.}
\centering
\setlength{\tabcolsep}{8pt}
\begin{tabular}{lcccc}
\toprule
& {BreastMNIST} 
& {RetinaMNIST} 
& {DermaMNIST} 
& {Average}\\

\midrule
End-to-End
& 85.26
& 55.25
& 81.23
& 73.91\\

Linear Probing
& 86.54
& 65.17
& 82.28
& 78.00\\

LoRA
& 88.25
& 66.83
& 87.43
& 80.84\\

BOFT
& 90.38
& 67.42
& 86.93
& 81.58\\

\midrule
\emph{MoPET}
& \underline{92.95}
& \underline{68.83}
& \underline{88.96}
& \underline{83.58}\\

\bottomrule
\end{tabular}
\label{tab:results3}
\end{table}

\subsection{Cross-Dataset Feature Sharing via Auxiliary Booster Datasets}
Finally, we investigate whether auxiliary datasets can further boost the predictive accuracy of individual target domains during \emph{MoPET}'s joint training. To evaluate this paradigm, we select three distinct booster configurations. Rather than adapting the DINOv3 backbone exclusively to a single dataset, we jointly train it alongside additional support datasets. Specifically, we co-train BreastMNIST with Blood-, Derma-, and PathMNIST; RetinaMNIST with Blood-, Breast-, Path-, and OrganAMNIST; and DermaMNIST with Blood-, OCT-, and OrganSMNIST.
\tablename~\ref{tab:results3} reports test accuracy on the three target datasets as the mean over three seeds. For reference, we include traditional fine-tuning and the isolated adapters. With its auxiliary pools, \emph{MoPET} reaches 92.95\% on Breast-, 68.83\% on Retina-, and 88.96\% on DermaMNIST, for an average of 83.58\%, ahead of every isolated baseline. The gains are largest on the smallest targets, Breast- and RetinaMNIST (546 and 1{,}080 training images), indicating that co-training with additional data benefits data-constrained domains the most.

\section{Discussion and Conclusion}
\label{sec:conclusion}
We introduce \emph{MoPET}, a mixture-of-experts method that couples a learned sparse router with low-rank experts so that a single frozen foundation model can serve many medical classification tasks at once. Across twelve datasets we first establish that parameter-efficient fine-tuning outperforms full end-to-end network updates, which motivates building the unified model from low-rank adapters. We then show that one \emph{MoPET} model consolidates four heterogeneous datasets and, on average, exceeds the best per-dataset adapter. Finally, co-training with auxiliary datasets improves accuracy on data-constrained targets, with the largest gains on the smallest datasets. The same joint formulation thus serves two ends, a single unified model or a boosted individual target.

\subsubsection{Limitations.} Several questions remain open. Key design choices, notably the expert count $K$, the top-$k$ activation, and the LoRA-to-BOFT ratio, are likely dataset dependent and warrant a more rigorous ablation than our scope allows. We also do not directly analyze the learned routing patterns; whether experts specialize by modality or anatomy, as intended, remains to be verified through an explicit study of expert utilization. Our cross-dataset and booster experiments further rely on hand-selected dataset subsets chosen to span modalities and scales; a systematic sweep over combinations, together with control conditions using randomly selected or unrelated auxiliary data, would be needed to separate semantically driven transfer from the effect of added training data. Finally, we do not benchmark against dedicated multi-task medical pretraining approaches, which would form an interesting comparison for the cross-dataset feature-sharing setting.

\begin{credits}
\subsubsection{\ackname} This study was funded through the Hightech Agenda Bayern (HTA) of the Free State of Bavaria, Germany.

\subsubsection{\discintname} The authors have no competing interests to declare that are relevant to the content of this article. 
\end{credits}
\bibliographystyle{splncs04}
\bibliography{mybibliography}




% LNCS Homepage, \url{http://www.springer.com/lncs}, last accessed 2023/10/25
% \end{thebibliography}

% --- arXiv-only supplementary material ---------------------------------
% Keep COMMENTED for the MICCAI/EMA proceedings submission (8+2 page limit).
% Uncomment to build the self-contained arXiv version with the appendix.
% \newpage
\input{supplementary}
% -----------------------------------------------------------------------

\end{document}

%% file: supplementary.tex
% ======================================================================
% Supplementary material (arXiv-only, self-contained).
% Not referenced from the main paper; mirrors the main-paper section order.
% Compile as part of the arXiv build after \end{document} content of the
% main file is replaced by an \input of this file, or append directly.
% ======================================================================
\clearpage
\appendix
\setcounter{page}{1}\renewcommand{\thepage}{S\arabic{page}}

% Hyperref anchor fixes for Sections & Tables
\renewcommand{\theHsection}{supp.\thesection}
\renewcommand{\theHtable}{supp.\thetable}

% Automatically reset and prefix tables per section (Table A1, B1...)
\numberwithin{table}{section}
\renewcommand{\thetable}{\thesection\arabic{table}}

\section*{Overview of Supplementary Material}
This supplementary material complements the main paper with additional
experimental detail and analyses. \textbf{Section~\ref{sec:experimentalDetails}} gives the full
experimental setup (compute, randomness, training configuration, preprocessing,
and per-method hyperparameters) needed to reproduce the reported statistics.
\textbf{Section~\ref{sec:extendedPeft}} extends the backbone/PEFT comparison with
per-backbone AUC and a per-dataset breakdown, and shows that the main-paper
ranking holds throughout: LoRA and BOFT are the strongest adapters on every
backbone, and their per-dataset complementarity motivates \emph{MoPET}'s hybrid expert
pool. \textbf{Section~\ref{sec:stats}} reports the complete Wilcoxon and Friedman
tests behind the significance claims, confirming that LoRA and BOFT are
statistically indistinguishable from each other yet separate from every other
method. \textbf{Section~\ref{sec:efficiency}} quantifies the parameter budget,
placing \emph{MoPET} at 8.7\% of the backbone while replacing a per-task collection of
adapters. \textbf{Section~\ref{sec:robustness}} evaluates MedMNIST-C corruptions
and finds the clean-data method ranking preserved, with AdaptFormer degrading
most.

% ======================================================================
\section{Experimental Details}
\label{sec:experimentalDetails}

\subsection{Computation}
All experiments were conducted on NVIDIA RTX A5000 and L40S GPUs. Backbones are
loaded from the \texttt{timm} library~\cite{rw2019timm} and kept frozen except
for end-to-end fine-tuning. Only the parameters introduced by each fine-tuning
method (and the task-specific classification heads) are updated.

\subsection{Randomness}
Each reported number is the mean over three independent training runs, each with
different random initialization, data ordering, and (for \emph{MoPET}) sampling. We
report the mean to characterize typical performance rather than a single seeded
outcome. The source code and run configurations are released to support reproduction of our full protocol and achieved results.

\subsection{Training Configuration}
Table~\ref{tab:supp_trainconfig} summarizes the optimization setup. The two
experiment families share optimizer, schedule, epoch budget, and early-stopping
criterion. They differ in batch size, where the unified \emph{MoPET} model uses a larger
effective batch realized through gradient accumulation. Learning rate and weight
decay correspond to the PyTorch AdamW defaults. For the DINOv3 linear-probing and
end-to-end baselines, the learning rate is reduced to $1\times10^{-4}$, following
the procedure of~\cite{Doerrich2025}.

\begin{table}[htb]
\centering
\caption{Training configuration. ``Isolated PEFT'' refers to the per-dataset
backbone/PEFT comparison (main paper Sec.~3.1). ``Unified \emph{MoPET}'' refers to the
cross-dataset and booster experiments (main paper Secs.~3.2--3.3).}
\label{tab:supp_trainconfig}
\setlength{\tabcolsep}{6pt}
\begin{tabular}{lll}
\toprule
Setting & Isolated PEFT & Unified \emph{MoPET} \\
\midrule
Optimizer            & AdamW~\cite{loshchilov2019decoupledweightdecayregularization} & AdamW \\
Learning rate        & $1\times10^{-3}$\,$^{\dagger}$ & $1\times10^{-3}$ \\
Weight decay         & $1\times10^{-2}$ & $1\times10^{-2}$ \\
LR schedule          & CosineAnnealingLR, one cycle & CosineAnnealingLR, one cycle \\
Max epochs           & 75 & 75 \\
Early stopping       & patience 10 (val.) & patience 10 (val.) \\
Batch size           & 64 & 128 (gradient accumulation) \\
Load-balancing $\lambda$ & --- & 0.05 \\
Loss                 & CE (BCE-with-logits, Chest) & CE $+\,\lambda\,\mathcal{L}_{\text{load}}$ \\
\bottomrule
\end{tabular}
\\[2pt]
{\footnotesize $^{\dagger}$ Reduced to $1\times10^{-4}$ for DINOv3 linear probing and end-to-end fine-tuning.}
\end{table}

\subsection{Preprocessing and Augmentation}
All datasets are taken from the MedMNIST+ collection at $224\times224$ pixels and
follow the official train/validation/test splits. Because the images are already
at $224\times224$, no padding is applied. For the DINOv3 backbone they are resized
to $256\times256$ with bicubic interpolation to match the backbone input
resolution. Inputs are normalized with the backbone's default
(ImageNet) channel statistics provided by \texttt{timm}. We apply \emph{no}
train-time data augmentation beyond padding, resizing, and normalization; the only
stochasticity in the input pipeline is the per-run random seed governing
initialization and sampling. Pretrained weights are loaded from \texttt{timm} and
the classification heads are randomly initialized. ChestMNIST is a multi-label
binary task and uses a binary cross-entropy-with-logits objective; all other
datasets use standard cross-entropy.

\subsection{Method Hyperparameters}
The isolated PEFT baselines use: LoRA~\cite{Hu2021} with rank $r=8$ and scaling
$\alpha=8$; BOFT~\cite{liu2024boft} with block size $8$ and butterfly factor $1$;
FourierFT~\cite{Gao2024FourierFT} with $n=1000$ frequencies and $\alpha=150$; and
AdaptFormer~\cite{chen2022adaptformer} with bottleneck dimension $64$ and scaling
$1$. All adapters are injected into the query, key, and value projections of every
Transformer block. \emph{MoPET} uses $K=32$ experts per block (20 LoRA experts with
$r=8,\alpha=8$ and 12 BOFT experts with block size $8$, butterfly factor $1$), a
shared frozen path, and a sparse router activating the top-$k=12$ experts per
forward pass, totalling $7.4$M trainable parameters ($\approx 8.7\%$ of the
$86$M backbone).

% ======================================================================
\section{Extended PEFT Comparison}
\label{sec:extendedPeft}
To confirm that the backbone and PEFT ranking of main-paper Table~1 does not depend
on the accuracy metric or on dataset-level averaging, we extend it along two axes:
we report the area under the ROC curve (AUC) alongside accuracy for every backbone,
and we decompose the DINOv3 accuracy into the individual datasets. Both are averaged over three seeds.

\subsubsection{Results}
Table~\ref{tab:supp_aggregate} shows the ranking is stable under AUC: LoRA and BOFT
are the strongest adapters on all three backbones, AdaptFormer is the weakest, and
DINOv3 is the strongest backbone for the high-performing methods, with the AUC
ordering mirroring the accuracy ordering.

\begin{table}[htb]
\centering
\caption{Mean test accuracy (ACC, \%) and AUC (\%) per fine-tuning method and
backbone, averaged over all twelve datasets and three random
seeds. Best ACC and AUC per backbone in bold. LoRA
attains the best accuracy on every backbone, and the AUC ordering mirrors the
accuracy ordering.}
\label{tab:supp_aggregate}
\setlength{\tabcolsep}{6pt}
\footnotesize
\begin{tabular}{lcccccccc}
\toprule
& & \multicolumn{3}{c}{ACC (\%)} & & \multicolumn{3}{c}{AUC (\%)} \\
\cmidrule(lr){3-5}\cmidrule(lr){7-9}
Method & & CLIP & DINO & DINOv3 & & CLIP & DINO & DINOv3 \\
\midrule
AdaptFormer & & $82.84$ & $82.08$ & $79.82$ & & 93.02 & 92.84 & 90.02 \\
BOFT        & & $87.37$ & $87.28$ & $88.62$ & & 95.84 & 95.90 & 96.19 \\
FourierFT   & & $84.57$ & $85.43$ & $86.72$ & & 94.79 & 95.04 & 95.18 \\
LoRA        & & $\mathbf{88.04}$ & $\mathbf{88.14}$ & $\mathbf{88.97}$ & & $\mathbf{95.99}$ & $\mathbf{96.14}$ & $\mathbf{96.24}$ \\
endToEnd    & & $82.75$ & $84.84$ & $86.50$ & & 91.83 & 93.90 & 94.10 \\
\bottomrule
\end{tabular}
\end{table}

Table~\ref{tab:supp_perdataset_dinov3} decomposes the DINOv3 average into the
individual datasets. No single adapter dominates: LoRA leads on most datasets while
among the adapters BOFT leads on Blood-, Breast-, Path-, Retina-, and TissueMNIST.
This per-dataset complementarity motivates \emph{MoPET}'s hybrid LoRA/BOFT expert pool
rather than a single-family pool.

\begin{table}[htb]
\centering
\caption{Per-dataset test accuracy (\%) for the DINOv3 backbone at resolution 256, as mean $\pm$ standard deviation over three seeds. Best method
per dataset in bold. No single adapter dominates: LoRA and BOFT alternate as the
strongest adapter across datasets, which motivates \emph{MoPET}'s hybrid pool.}
\label{tab:supp_perdataset_dinov3}
\setlength{\tabcolsep}{4pt}\footnotesize
\resizebox{\textwidth}{!}{%
\begin{tabular}{lccccc}
\toprule
Dataset & LoRA & BOFT & FourierFT & AdaptFormer & end-to-end \\
\midrule
Blood     & $98.75{\scriptstyle\pm0.27}$ & $\mathbf{98.81}{\scriptstyle\pm0.10}$ & $98.40{\scriptstyle\pm0.06}$ & $96.50{\scriptstyle\pm0.17}$ & $98.49{\scriptstyle\pm0.44}$ \\
Breast    & $88.25{\scriptstyle\pm0.74}$ & $\mathbf{90.38}{\scriptstyle\pm0.64}$ & $88.46{\scriptstyle\pm0.64}$ & $73.29{\scriptstyle\pm2.06}$ & $85.26{\scriptstyle\pm3.57}$ \\
Chest     & $\mathbf{94.83}{\scriptstyle\pm0.02}$ & $94.81{\scriptstyle\pm0.05}$ & $94.77{\scriptstyle\pm0.00}$ & $94.75{\scriptstyle\pm0.01}$ & $94.79{\scriptstyle\pm0.01}$ \\
Derma     & $\mathbf{87.43}{\scriptstyle\pm1.35}$ & $86.93{\scriptstyle\pm1.14}$ & $83.18{\scriptstyle\pm0.33}$ & $69.89{\scriptstyle\pm0.26}$ & $81.23{\scriptstyle\pm1.25}$ \\
OCT       & $\mathbf{91.87}{\scriptstyle\pm1.84}$ & $90.10{\scriptstyle\pm2.45}$ & $88.13{\scriptstyle\pm0.38}$ & $71.47{\scriptstyle\pm5.42}$ & $90.30{\scriptstyle\pm1.99}$ \\
OrganA    & $\mathbf{97.72}{\scriptstyle\pm0.09}$ & $97.09{\scriptstyle\pm0.16}$ & $95.48{\scriptstyle\pm0.08}$ & $92.75{\scriptstyle\pm1.47}$ & $97.49{\scriptstyle\pm0.15}$ \\
OrganC    & $95.16{\scriptstyle\pm0.67}$ & $93.80{\scriptstyle\pm0.29}$ & $92.69{\scriptstyle\pm0.11}$ & $88.10{\scriptstyle\pm1.00}$ & $\mathbf{95.24}{\scriptstyle\pm0.17}$ \\
OrganS    & $83.29{\scriptstyle\pm0.07}$ & $81.47{\scriptstyle\pm0.83}$ & $79.97{\scriptstyle\pm0.21}$ & $73.97{\scriptstyle\pm0.36}$ & $\mathbf{84.33}{\scriptstyle\pm0.31}$ \\
Path      & $95.07{\scriptstyle\pm0.50}$ & $95.21{\scriptstyle\pm0.78}$ & $94.52{\scriptstyle\pm0.16}$ & $89.06{\scriptstyle\pm1.45}$ & $\mathbf{96.13}{\scriptstyle\pm0.30}$ \\
Pneumonia & $\mathbf{93.38}{\scriptstyle\pm0.46}$ & $91.99{\scriptstyle\pm1.25}$ & $90.76{\scriptstyle\pm0.19}$ & $83.74{\scriptstyle\pm0.28}$ & $87.23{\scriptstyle\pm0.91}$ \\
Retina    & $66.83{\scriptstyle\pm2.43}$ & $\mathbf{67.42}{\scriptstyle\pm0.29}$ & $65.50{\scriptstyle\pm0.25}$ & $54.53{\scriptstyle\pm0.76}$ & $55.25{\scriptstyle\pm2.88}$ \\
Tissue    & $75.08{\scriptstyle\pm0.34}$ & $\mathbf{75.38}{\scriptstyle\pm0.13}$ & $68.77{\scriptstyle\pm0.08}$ & $69.74{\scriptstyle\pm0.34}$ & $72.23{\scriptstyle\pm1.84}$ \\
\midrule
Avg. & $\mathbf{88.97}{\scriptstyle\pm0.73}$ & $88.62{\scriptstyle\pm0.68}$ & $86.72{\scriptstyle\pm0.21}$ & $79.82{\scriptstyle\pm1.13}$ & $86.50{\scriptstyle\pm1.15}$ \\
\bottomrule
\end{tabular}}
\end{table}

% ======================================================================
\section{Statistical Testing}
\label{sec:stats}
We report the complete pairwise Wilcoxon signed-rank tests with Bonferroni
correction and the corresponding Friedman omnibus statistics that underlie the
significance statements in the main paper. All tests use $\alpha=0.05$. The method and backbone comparisons are computed over three seeds for each fine-tuning method, backbone, and dataset. Reported entries are Bonferroni-corrected $p$-values.

\subsection{Fine-Tuning Methods}
In the main paper, LoRA and BOFT significantly outperform full fine-tuning and the weaker PEFT adapters (Section~3.1). We establish this with an
omnibus Friedman test across methods, followed by pairwise Wilcoxon signed-rank
tests with Bonferroni correction over the twelve datasets.

\subsubsection{Results}
A Friedman test across methods yields $\chi^2 = 104.85$, $p = 5.0\times10^{-21}$.
LoRA and BOFT each significantly outperform end-to-end fine-tuning, FourierFT,
linear probing, and AdaptFormer, but are statistically indistinguishable from one
another ($p = 0.077$). End-to-end fine-tuning is indistinguishable from FourierFT
and linear probing ($p = 1.00$) (Table~\ref{tab:supp_wilcoxon_methods}).

\begin{table}[htb]
\centering
\caption{Pairwise Wilcoxon signed-rank tests (Bonferroni-corrected $p$-values)
for fine-tuning methods across all twelve datasets. Friedman
$\chi^2 = 104.85$, $p = 5.0\times10^{-21}$. LoRA and BOFT are mutually
indistinguishable yet each separates from every other method.}
\label{tab:supp_wilcoxon_methods}
\setlength{\tabcolsep}{4pt}\footnotesize
\resizebox{\textwidth}{!}{%
\begin{tabular}{lcccccc}
\toprule
 & LoRA & BOFT & endToEnd & FourierFT & linProbe & AdaptFormer \\
\midrule
LoRA        & --- & $7.75{\times}10^{-2}$ & $4.67{\times}10^{-7}$ & $4.67{\times}10^{-7}$ & $9.57{\times}10^{-6}$ & $4.37{\times}10^{-10}$ \\
BOFT        & $7.75{\times}10^{-2}$ & --- & $1.21{\times}10^{-3}$ & $3.66{\times}10^{-5}$ & $1.87{\times}10^{-5}$ & $3.06{\times}10^{-9}$ \\
endToEnd    & $4.67{\times}10^{-7}$ & $1.21{\times}10^{-3}$ & --- & $1.00$ & $1.00$ & $3.44{\times}10^{-3}$ \\
FourierFT   & $4.67{\times}10^{-7}$ & $3.66{\times}10^{-5}$ & $1.00$ & --- & $9.98{\times}10^{-4}$ & $9.93{\times}10^{-4}$ \\
linProbe    & $9.57{\times}10^{-6}$ & $1.87{\times}10^{-5}$ & $1.00$ & $9.98{\times}10^{-4}$ & --- & $2.63{\times}10^{-1}$ \\
AdaptFormer & $4.37{\times}10^{-10}$ & $3.06{\times}10^{-9}$ & $3.44{\times}10^{-3}$ & $9.93{\times}10^{-4}$ & $2.63{\times}10^{-1}$ & --- \\
\bottomrule
\end{tabular}}
\end{table}

\subsection{Foundation Models}
In the main paper, the fine-tuning method matters far more than the backbone
(Section~3.1). We quantify the backbone effect here, first across the
parameter-efficient methods alone and then including end-to-end fine-tuning and
linear probing, using the same Friedman and pairwise Wilcoxon tests.

\subsubsection{Results}
Under parameter-efficient settings the three backbones are statistically
indistinguishable (Friedman $\chi^2 = 3.63$, $p = 0.163$). When the comparison
additionally includes end-to-end fine-tuning and linear probing, the backbones do
differ (Friedman $\chi^2 = 16.23$, $p = 3.0\times10^{-4}$): the pairwise tests in
Table~\ref{tab:supp_wilcoxon_models} separate CLIP from the two DINO variants,
while DINO and DINOv3 remain indistinguishable.

\begin{table}[htb]
\centering
\caption{Pairwise Wilcoxon signed-rank tests (Bonferroni-corrected $p$-values)
for foundation models on clean data, across all fine-tuning methods and datasets. Friedman $\chi^2 = 16.23$, $p = 3.0\times10^{-4}$. CLIP separates from both DINO variants, which are themselves indistinguishable.}
\label{tab:supp_wilcoxon_models}
\setlength{\tabcolsep}{6pt}\footnotesize
\begin{tabular}{lccc}
\toprule
 & DINOv3 & DINO & CLIP \\
\midrule
DINOv3 & --- & $0.188$ & $1.05{\times}10^{-4}$ \\
DINO   & $0.188$ & --- & $7.00{\times}10^{-3}$ \\
CLIP   & $1.05{\times}10^{-4}$ & $7.00{\times}10^{-3}$ & --- \\
\bottomrule
\end{tabular}
\end{table}

\subsection{Effect of Input Resolution}
To justify the single $224\times224$ resolution used throughout the main paper, we
ablate the input resolution. We compare $128\times128$ against $224\times224$ under
otherwise identical configurations, reporting paired accuracy gains and Wilcoxon
signed-rank tests per fine-tuning method and per backbone.

\subsubsection{Results}
Increasing the resolution from $128\times128$ to $224\times224$ yields small but,
for most methods and backbones, statistically significant accuracy gains
(Tables~\ref{tab:supp_res_methods} and~\ref{tab:supp_res_models}). The gain is
positive for every method and significant for all but AdaptFormer, and every
backbone improves, DINOv3 the most (mean $+1.46$ points). The gains are modest,
supporting the single $224$ resolution in the main experiments rather than a
multi-resolution protocol.

\begin{table}[htb]
\centering
\caption{Wilcoxon signed-rank test for the effect of increased input resolution ($224\times224$ vs.\ $128\times128$),
per fine-tuning method. Gains are paired $224$-minus-$128$ differences. The gain
is positive for every method and significant for all but AdaptFormer.}
\label{tab:supp_res_methods}
\setlength{\tabcolsep}{6pt}\footnotesize
\resizebox{\textwidth}{!}{%
\begin{tabular}{lccccc}
\toprule
Method & $n$ pairs & Median gain (\%) & Mean gain (\%) & Wilcoxon stat. & $p$-value \\
\midrule
AdaptFormer   & 36 & 0.218 & 0.419 & 269.0 & $0.323$ \\
BOFT          & 36 & 0.308 & 0.856 & 140.0 & $1.85{\times}10^{-3}$ \\
FourierFT     & 36 & 0.706 & 1.231 & 71.0  & $9.00{\times}10^{-6}$ \\
LoRA          & 36 & 0.349 & 0.970 & 176.0 & $1.27{\times}10^{-2}$ \\
endToEnd      & 36 & 0.203 & 0.683 & 206.0 & $4.60{\times}10^{-2}$ \\
linearProbing & 36 & 0.590 & 1.120 & 101.0 & $1.26{\times}10^{-4}$ \\
\bottomrule
\end{tabular}}
\end{table}

\begin{table}[htb]
\centering
\caption{Wilcoxon signed-rank test for the effect of increased input resolution,
per foundation model. Every backbone improves with resolution, DINOv3 the most
(mean $+1.46$ points).}
\label{tab:supp_res_models}
\setlength{\tabcolsep}{6pt}\footnotesize
\resizebox{\textwidth}{!}{%
\begin{tabular}{lccccc}
\toprule
Architecture & $n$ pairs & Median gain (\%) & Mean gain (\%) & Wilcoxon stat. & $p$-value \\
\midrule
CLIP   & 72 & 0.200 & 0.623 & 738.0 & $1.23{\times}10^{-3}$ \\
DINO   & 72 & 0.258 & 0.557 & 842.5 & $8.15{\times}10^{-3}$ \\
DINOv3 & 72 & 0.983 & 1.460 & 374.0 & $1.33{\times}10^{-7}$ \\
\bottomrule
\end{tabular}}
\end{table}

% ======================================================================
\section{Efficiency Trade-off}
\label{sec:efficiency}
To quantify \emph{MoPET}'s cost against the isolated adapters, we report the
trainable-parameter budget of each method for a single MedMNIST task and compare it
against \emph{MoPET}, which unifies all tasks in one model (Section~3.2).

\subsubsection{Results}
Table~\ref{tab:supp_params} reports the budget, which is essentially identical
across the three backbones as they share the same Base architecture. A single
isolated adapter is far lighter than \emph{MoPET} (e.g.\ LoRA $0.30$M vs.\ $7.4$M) but
serves only one task. At $7.4$M parameters ($8.7\%$ of the backbone), \emph{MoPET}
replaces a per-task collection of adapters with one model.

\begin{table}[htb]
\centering
\caption{Trainable parameters per method (millions and percentage of the $86$M
backbone). \emph{MoPET} counts the experts, router, and the per-dataset heads.}
\label{tab:supp_params}
\setlength{\tabcolsep}{8pt}
\begin{tabular}{lcc}
\toprule
Method & Trainable params (M) & \% of backbone \\
\midrule
Full (end-to-end) & 86      & 100\% \\
AdaptFormer       & 1.209   & 1.4\% \\
LoRA              & 0.296   & 0.34\% \\
BOFT              & 0.103   & 0.12\% \\
FourierFT         & 0.0014  & 0.016\% \\
\midrule
\emph{MoPET}      & 7.4     & 8.7\% \\
\bottomrule
\end{tabular}
\end{table}

% ======================================================================
\section{Robustness on Corrupted Data}
\label{sec:robustness}
Clinical images are frequently corrupted at acquisition, so a deployable method must
retain its ranking under corruption. To test this, we train on clean data and
evaluate on the MedMNIST-C~\cite{DiSalvo2024} corrupted test sets. Each test image
receives one randomly drawn corruption (including the identity) at a random severity
within the dataset-specific range. To keep the comparison fair, the random choice of
corruption and severity is fixed per seed across methods, and the corruption API
operates at $224\times224$ resolution.

\subsubsection{Results}

Table~\ref{tab:supp_corrupt_drop} reports the per-method, per-backbone accuracy
drop from clean to corrupted data. The relative ordering of methods is preserved
under corruption: LoRA and BOFT
remain the most robust, end-to-end fine-tuning and FourierFT are intermediate, and
AdaptFormer degrades most (median accuracy drop of $19.3$ percentage points). At
the level of method/backbone pairings, the AdaptFormer--DINOv3 combination suffers
the largest degradation ($29.91$ percentage points), whereas end-to-end
fine-tuning with DINOv3 is the most robust pairing ($4.52$ points), consistent
with DINOv3 being the strongest clean-data backbone. Across backbones, median
accuracy decreases by roughly $8$ percentage points while the relative ranking
(DINOv3 $>$ DINO $>$ CLIP) is preserved.

\begin{table}[htb]
\centering
\caption{Accuracy drop (percentage points) from clean to MedMNIST-C corrupted test
data, per fine-tuning method and backbone (lower is more robust). Lowest drop per
backbone in bold. LoRA and BOFT are the most robust across backbones, AdaptFormer
degrades most, and the AdaptFormer--DINOv3 pairing is the least robust overall.}
\label{tab:supp_corrupt_drop}
\setlength{\tabcolsep}{6pt}\footnotesize
\begin{tabular}{lccccc}
\toprule
Architecture & AdaptFormer & BOFT & FourierFT & LoRA & end-to-end \\
\midrule
CLIP   & 14.83 & 7.08 & 9.65 & \textbf{6.19} & 7.72 \\
DINO   & 19.94 & 6.77 & 8.20 & \textbf{6.14} & 7.12 \\
DINOv3 & 29.91 & 6.51 & 6.59 & 5.02 & \textbf{4.52} \\
\bottomrule
\end{tabular}
\end{table}

Tables~\ref{tab:supp_corrupt_methods} and~\ref{tab:supp_corrupt_models} give the
pairwise tests on corrupted data. Across methods the differences are highly
significant (Friedman $\chi^2 = 77.17$, $p = 6.92\times10^{-16}$): LoRA differs
significantly from all other methods and BOFT from all except LoRA, while
end-to-end and FourierFT are indistinguishable. Across backbones the omnibus test
is only marginally significant ($\chi^2 = 6.18$, $p = 0.046$) and no pairwise
comparison survives correction, indicating that the choice of backbone has limited
impact on corrupted-data performance.

\begin{table}[htb]
\centering
\caption{Pairwise Wilcoxon signed-rank tests (Bonferroni-corrected $p$-values) for
fine-tuning methods on corrupted data. Friedman $\chi^2 = 77.17$,
$p = 6.92\times10^{-16}$. LoRA is the most robust method and separates from all
others, while end-to-end and FourierFT remain indistinguishable.}
\label{tab:supp_corrupt_methods}
\setlength{\tabcolsep}{4pt}\footnotesize
\begin{tabular}{lccccc}
\toprule
 & AdaptFormer & BOFT & FourierFT & LoRA & end-to-end \\
\midrule
AdaptFormer & --- & $2.98{\times}10^{-7}$ & $4.17{\times}10^{-6}$ & $2.98{\times}10^{-7}$ & $2.35{\times}10^{-4}$ \\
BOFT        & $2.98{\times}10^{-7}$ & --- & $3.28{\times}10^{-5}$ & $8.35{\times}10^{-3}$ & $2.47{\times}10^{-2}$ \\
FourierFT   & $4.17{\times}10^{-6}$ & $3.28{\times}10^{-5}$ & --- & $9.83{\times}10^{-6}$ & $1.00$ \\
LoRA        & $2.98{\times}10^{-7}$ & $8.35{\times}10^{-3}$ & $9.83{\times}10^{-6}$ & --- & $4.17{\times}10^{-6}$ \\
end-to-end  & $2.35{\times}10^{-4}$ & $2.47{\times}10^{-2}$ & $1.00$ & $4.17{\times}10^{-6}$ & --- \\
\bottomrule
\end{tabular}
\end{table}

\begin{table}[htb]
\centering
\caption{Pairwise Wilcoxon signed-rank tests (Bonferroni-corrected $p$-values) for
foundation models on corrupted data. Friedman $\chi^2 = 6.18$, $p = 0.046$. No
pair of backbones separates after correction, so backbone choice barely affects
corrupted-data accuracy.}
\label{tab:supp_corrupt_models}
\setlength{\tabcolsep}{6pt}\footnotesize
\begin{tabular}{lccc}
\toprule
 & DINOv3 & DINO & CLIP \\
\midrule
DINOv3 & --- & $0.631$ & $0.368$ \\
DINO   & $0.631$ & --- & $1.000$ \\
CLIP   & $0.368$ & $1.000$ & --- \\
\bottomrule
\end{tabular}
\end{table}